\documentclass[usenatbib]{mn2e}

\usepackage{graphicx}
\usepackage{epsfig}
\usepackage{amsmath}
\usepackage{natbib}
\usepackage{times}

\newcommand{\solarmass}{$h^{-1}{\rm M_\odot}$}
\newcommand{\kpch}{$h^{-1}{\rm kpc}$}

\newcommand{\gtsima}{\mbox{$\; \buildrel > \over \sim \;$}} 
 
\newcommand{\simgt}{\lower.5ex\hbox{\gtsima}}            % > over ~

\def\gsim { \lower .75ex \hbox{$\sim$} \llap{\raise .27ex \hbox{$>$}}}
\def\lsim { \lower .75ex \hbox{$\sim$} \llap{\raise .27ex \hbox{$<$}}}
\title
{Star forming filaments in warm dark matter models}
\author[Gao, Theuns  \&Springel]
{Liang Gao$^{1,2}$\thanks{lgao@bao.ac.cn}, Tom Theuns$^{2,3}$ and Volker Springel$^{5,6}$\\
$^1$Key laboratory for Computational Astrophysics, Partner Group of
  the Max Planck Institute for Astrophysics, \\National Astronomical
  Observatories, Chinese Academy of Sciences, Beijing, 100012, China
  \\
$^2$Institute for Computational Cosmology, Department of Physics,
Durham University, South Road, Durham DH1 3LE, UK\\
$^3$ Department of Physics, University of Antwerp, Campus Groenenborger, Groenenborgerlaan 171, B-2020 Antwerpen, Belgium\\
$^5$Heidelberger Institut f\"{u}r Theoretische Studien, Schloss-Wolfsbrunnenweg 35, 69118 Heidelberg, Germany\\
$^6$Zentrum f\"ur Astronomie der Universit\"at Heidelberg, Astronomisches
Recheninstitut, M\"{o}nchhofstr. 12-14, 69120 Heidelberg, Germany\\
}

\begin{document}

\maketitle
\begin{abstract}
We performed a hydrodynamical cosmological simulation of the formation
of a Milky Way-like galaxy in a warm dark matter (WDM)
cosmology. Smooth and dense filaments, several co-moving mega parsec
long, form generically above $z\sim 2$ in this model. Atomic line
cooling allows gas in the centres of these filaments to cool to the
base of the cooling function, resulting in a very striking pattern of
extended Lyman-limit systems (LLSs). Observations of the correlation
function of LLSs might hence provide useful limits on the nature of
the dark matter. We argue that the self-shielding of filaments may
lead to a thermal instability resulting in star formation. We
implement a sub-grid model for this, and find that filaments rather
than haloes dominate star formation until $z\sim 6$ although this
depends on how stars form in WDM. Reionisation decreases the gas
density in filaments, and the more usual star formation in haloes
dominates below $z\sim 6$, although star formation in filaments
continues until $z=2$. Fifteen per cent of the stars of the $z=0$
galaxy formed in filaments. At higher redshift, these stars give
galaxies a stringy appearance, which, if observed, might be a strong
indication that the dark matter is warm. 
\end{abstract}


\section{Introduction} \label{sect:intro}
The $\Lambda$ cold dark matter (CDM) model  has had tremendous success
in providing a cosmogony that links the state of the very
high-redshift Universe as inferred from cosmic microwave background
observations and Big Bang nucleosynthesis considerations, with the
observed large-scale distribution of galaxies and its evolution at
later times. In this paradigm, the observed structures grew due to
gravitational amplification of initially small perturbations, perhaps
seeded by inflation \citep{Guth81}, with most of the gravitating mass
in the form of \lq dark matter\rq, an as of yet unknown type of matter
that apart from gravitationally, interacts only very weakly if at all
with baryonic matter and radiation, see e.g. \cite{FrenkWhite12} for a
recent review. Observations on smaller, sub galactic scales, have
proven more problematic for $\Lambda$CDM, with suggestions that the
low abundance \citep[e.g][]{Klypin99, Moore99} and shallow density
profiles \citep[e.g][]{Gilmore2007, de_vega2010} inferred for haloes
hosting Milky Way satellites are inconsistent with the numerous
substructures with cuspy density profiles that form in Milky Way-like
CDM haloes. Indeed the substructures in the haloes of the Aquarius project
\citep{Springel08b} and the GHALO project \citep{Stadel09} may be
difficult to reconcile with those inferred to host Milky Way
satellites. Even if many of the smaller DM substructures may not have
a sufficiently deep potential well to form stars after reionization
\citep[e.g][] {Efstathiou92, Benson02} and so may well be dark, some
of the more massive ones are probably too big to be affected
\citep{Okamoto08} and would hence appear to be \lq too big to
fail\rq\ \citep{Boylan-Kolchin11,Boylan-Kolchin12}. However the jury
on this is still out: the Milky Way's halo may simply be of lower mass
than those of the Aquarius haloes and hence have fewer massive
satellites \citep{Wang12}. The status of the {\em density} profiles of
the satellite's haloes - cored versus cuspy - is similarly
unresolved. \cite{Strigari10} claim that the stellar dynamics
observations of the satellites do not imply cores at all and hence may
be consistent with CDM cusps. But even if the satellites had cored
profile, this might result from the action of baryonic feedback
processes \citep{Pontzen12, Governato12}, and hence still be
consistent with CDM.

Irrespective of these astrophysical considerations, particle physics
has several candidates for the dark matter. A popular kind is a super
symmetric particle  with negligible intrinsic velocity dispersion
\citep{Bertone2005}, which would qualify as \lq cold\rq\ DM.  However
other viable candidates have considerable intrinsic velocities, and
these would constitute \lq warm\rq\ dark matter (WDM), for example a
keV scale gravitino or a sterile neutrino
\citep[e.g.][]{Dodelson94}. Such intrinsic velocities - as opposed to
velocities induced by gravity - have two (related) effects on the
formation of structure. Firstly, the motion of WDM particles quenches
the growth of structure below a \lq free-streaming scale\rq, which we
loosely define here as the the maximum distance over which such a
particle can travel. As a consequence, WDM haloes have far less
substructures with masses below the corresponding free-streaming mass
- potentially alleviating the \lq missing satellites\rq\ problem
discussed before \citep{Bode01}. Secondly, the intrinsic velocities of
WDM particles imply a finite phase-space density - which is conserved
during halo growth \citep[e.g.][]{Tremaine79, Hogan00}. This finite
phase space density might be the origin of the cores inferred to exist
in Milky Way satellite haloes - if indeed they are cored.

Even though these considerations stimulated much of the astronomical
interest in WDM, detailed analyses yielded slightly disappointing
results. \cite{Shao13} demonstrated that - although WDM haloes are
indeed cored - cores as large as suggested by Milky Way satellite
observations will only form if the WDM free-steaming lengths is so
large that the satellites themselves would fail to form, resulting in
a a \lq too small to succeed\rq\ problem (see also \cite{Maccio13}).
\cite{Lovell12} demonstrated that WDM does help with the satellites'
profiles: since the haloes hosting satellites form later in WDM than
in CDM on average, they tend to have lower central densities -
alleviating the  \lq too big to fail\rq\ problem. Requiring that
enough satellites form leads to a conservative lower limit to the
(thermal equivalent) WDM particle mass of ~1.5~keV \citep{Lovell13},
not far from the lower limits inferred from the Ly$\alpha$ forest by
\cite{Boyarsky09}. Consequently there is still a window for WDM to
have some effect on galaxy formation, although it seems that by itself
it will not resolve the issues with the \lq too big to
fail\rq\ problem, and the alleged presence of cores in satellite
haloes.

The study of \cite{Gao07b} points out that the formation of the first
stars could proceed very differently in a WDM Universe. The reason is
that the filaments that form naturally in hierarchical models have
potential wells that are so deep that gas in them forms molecular hydrogen 
and cools. Such filaments also form in CDM, yet
there they break-up into star forming mini haloes due to CDM's
small-scale power. Numerical fragmentation who's origin is discussed
by \cite{Wang07} prevented \cite{Gao07b} from making firm predictions
about the nature of first star formation in WDM, but they argue that
star formation may be very efficient and result in a range of stellar
masses, which is quite different from the inefficient formation of a
\lq single\rq\ massive star expected to form in CDM mini haloes
\citep[e.g.][]{Abel02}.

Here we present results from zoomed cosmological hydrodynamical
simulations that follow the formation of a Milky Way-like galaxy in
WDM. In particular we want to test whether the filaments that form
around such objects could host star formation initiated by {\em
  atomic} line cooling, which would be the galaxy formation analogue
of first star formation in WDM discussed by \cite{Gao07b}. We
introduce the simulations in Section~2 and present our results in
Section~3, Section~4 summarises.

%\section{Zoomed simulations of the formation of a galaxy in WDM:
%set-up}
\section{Zoomed simulations of the formation of a galaxy in WDM: set-up}
Our simulations were performed with {\sc gadget-3}, a Tree plus
smoothed particle hydrodynamics (SPH,  \cite{Lucy77, Gingold77}, see
e.g. \cite{Springel10} for a recent review) code based on {\sc
  gadget-2} last described by \cite{Springel05}. We include gas
cooling and photo-heating of optically thin primordial gas in the
presence of an imposed UV/X-ray background from \cite{Haardt96}, under
the assumption of collisional ionization equilibrium as in
\cite{Katz96}.  We assume H{\sc I} reionization occurs at redshift
$z=6$. We don't include a model for molecular hydrogen formation and
hence neglect effects of \lq first stars\rq. The halo we simulate is
\lq halo A\rq\ from the Aquarius project \citep{Springel08b}, and the
assumed cosmological parameters ($\Omega_m=0.25$,
$\Omega_{\Lambda}=0.75$, $\sigma_8=0.9$ and $h=0.73$) are taken from
that paper. 

Using a zoomed technique, the formation of structure in a box of
100$h^{-1}$ co-moving mega parsecs on a side is followed by massive
dark matter particles, with the proto-halo itself represented by much
lower mass particles with masses of $2.4 \times 10^5$\solarmass\ and
$5.2 \times 10^4$\solarmass, for dark matter and gas, respectively.

\begin{table}
\begin{tabular}{lcc}
\hline
$M_{\rm DM}$ [\solarmass] & $M_{\rm GAS}$ [\solarmass] & $\epsilon$ [\kpch] \\
\hline
$2.35 \times 10^5$ & $5.16 \times 10^4$ & $0.5$ \\
\hline
\end{tabular}
\caption{Basic parameters of our numerical simulation. $M_{\rm DM}$ is
the dark matter particle mass, $M_{\rm GAS}$ the SPH particle mass;
$\epsilon$ is the co-moving softening length.}
\end{table}

\begin{figure*}
\hspace{0.13cm}\resizebox{16cm}{!}{\includegraphics{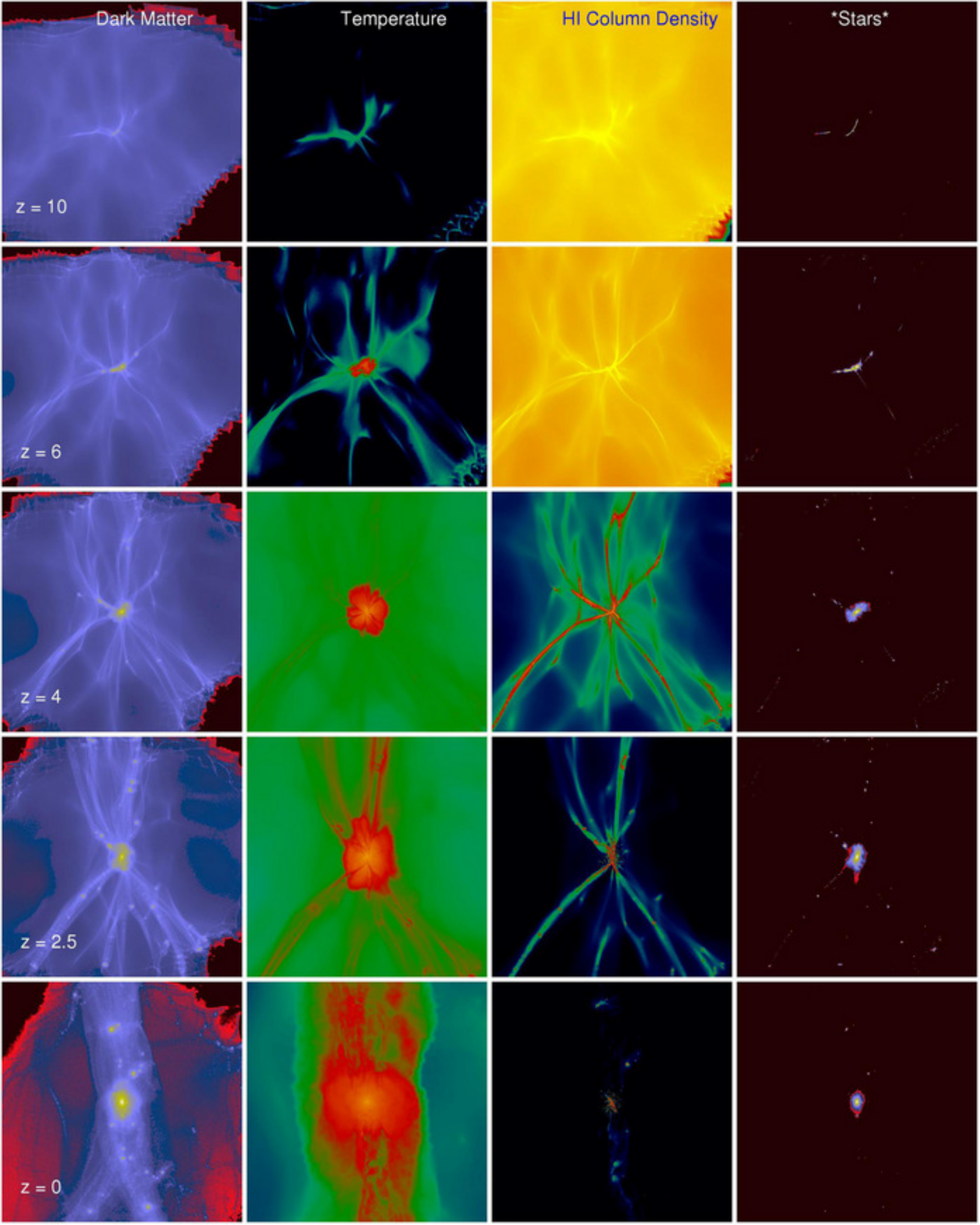}}
\hspace{0.13cm}\resizebox{3.85cm}{!}{\includegraphics{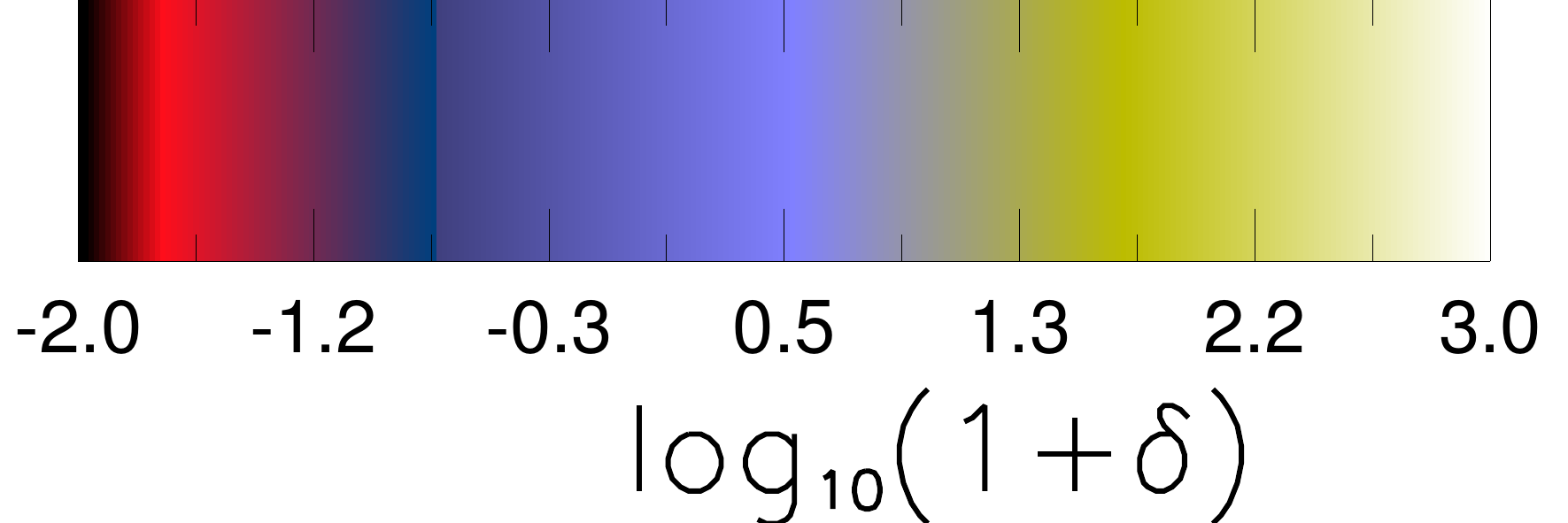}}
\hspace{0.13cm}\resizebox{3.85cm}{!}{\includegraphics{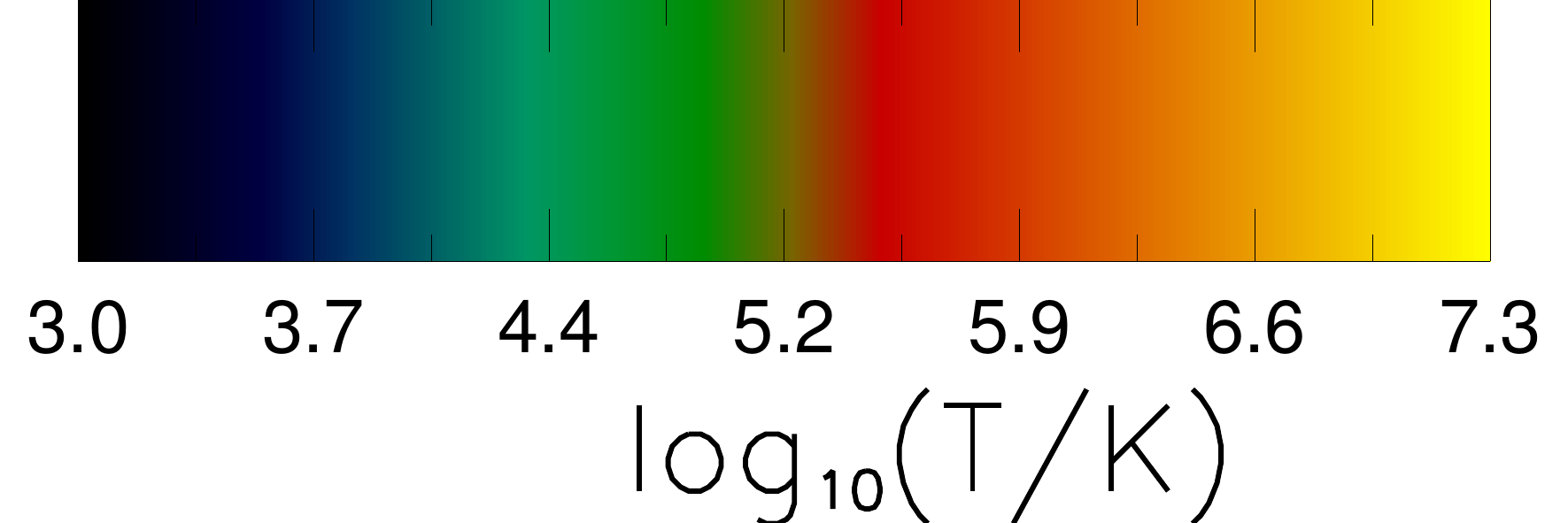}}
\hspace{0.13cm}\resizebox{3.85cm}{!}{\includegraphics{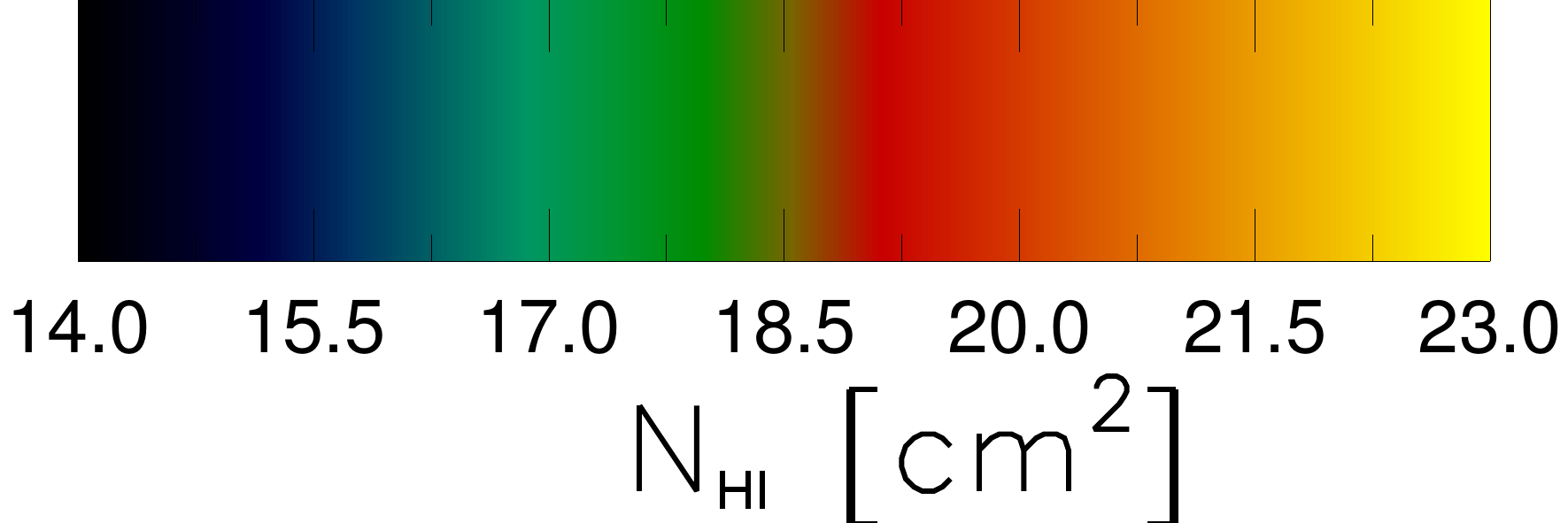}}
\hspace{0.13cm}\resizebox{3.85cm}{!}{\includegraphics{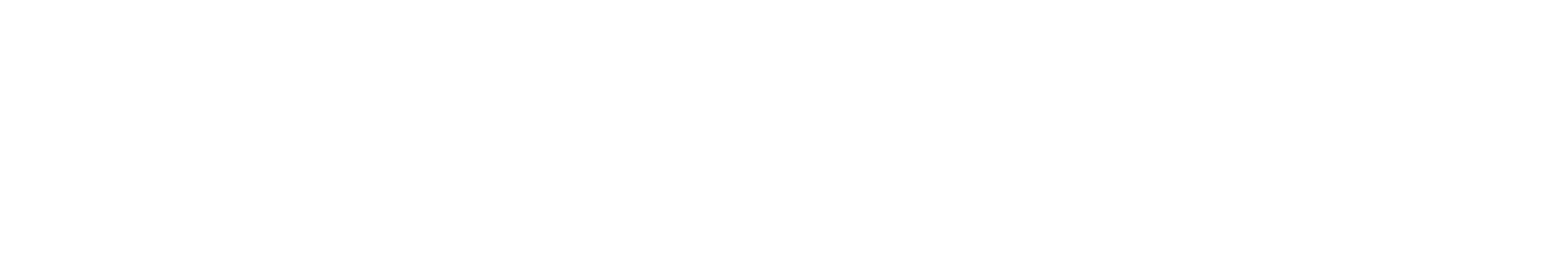}}
\caption{Image of the galaxy forming in the Aquarius-A simulation in a
  warm dark matter Universe, shown at various redshifts (different
  rows, redshifts are indicated in the left panels). From left to
  right, different panels indicate the dark matter over density, gas
  temperature, hydrogen column density, and the density of stars,
  respectively; each panel is 4 co-moving $h^{-1}$Mpc across. The
  panels are centred on the galaxy, and colour bars are presented for
  the left three columns. The big filaments visible in dark matter and
  gas are very striking, as time progresses they become fewer in
  number but continue to dominate the visual impression. Artificial
  fragmentation is visible in the dark matter, but is less pronounced
  in the gas. Gas shock heats into the filaments, partially cools, and
  drains into the main galaxy. The density map of stars (right column)
  is normalised separately for each redshift, to bring-out the
  formation of stars in the filaments.}
\label{Fig1}
\end{figure*}

The WDM linear power-spectrum is calculated by exponentially
truncating the corresponding CDM power below a free-streaming scale
that corresponds to that of a $1.5$~keV equivalent thermal relic
particle. Such an exponential truncation of the power spectrum does
not do justice to the real underlying physics of WDM, but the effects
on galaxy formation discussed here will nevertheless be quite
generic. We use the same random phases for initialising the Gaussian
field as for Aq-A, allowing a straightforward comparison between the
CDM and WDM runs. In order to illustrate differences in CDM and
  WDM model, we have repeated the simulation for the case of cold dark
  matter.

The main objective of this exploratory study is to investigate whether
in a WDM model star formation is likely to occur in filaments around
young galaxies. Unfortunately it is unclear which physical conditions
are required for this. In {\em galaxies}, stars form in molecular
clouds which themselves are thought to result from disk
instabilities. A prerequisite for this to happen is that gas
self-shields from the UV background allowing it to cool radiatively,
as in the model of \cite{Schaye04}. We implement this by specifying a
density threshold above which gas (i.e. SPH) particles are converted
to stars. For this to happen we require that the physical hydrogen
density $n_{\rm H}>0.1~{\rm cm}^{-3}$ and that the over density
$\rho/\bar \rho>2000$\footnote{We have used a higher denisty threshold
  $n_{\rm H}>5~{\rm cm}^{-3}$ for the star formation, and re-run our
  simulation to redshift $6$, the results are qualitatively similar to
what we present below {\bf (see appendix)}.}. Hydrodynamical simulations with radiative
transfer applied in post-processing \citep[e.g.][]{Altay11,Rahmati13}
show that cosmic gas turns mostly neutral above such a
threshold. These radiative transfer calculations apply to CDM - but
the same physics should of course apply to the WDM case as well.  A
more physical description should consider the instability of the
(nearly) 1 dimensional filament \citep{Inutsuka97}, but our simulation
lacks the physics and resolution to warrant a more detailed star
formation description. For the same reason, we neglect feedback
  processes in  this study. 

In CDM galaxy formation simulations, the rate at which galaxies form
stars is in fact mostly set by their feedback efficiency rather than
by the assumed star formation rate itself, as explicitly demonstrated
using the {\sc owls} simulations by \cite{Haas13a, Haas13b}. The
reason for this is that star formation is self-regulating, at least in these
models \citep{Schaye10}. In the present calculations we do not include
any feedback from forming stars, and that is of course a concern. Our
motivation is that we wish to examine whether stars form at all. In
addition, the feedback descriptions in use are tuned to CDM, and may
need to be modified before being applied to star formation in
filaments.
%Also, feedback may be much less efficient in filaments than
%in disks, as hot bubbles produced by super nova explosion that are
%thought to drive feedback can reach low densities much easier in a
%filament than in a disk. For example the simulations in
%\cite{Theuns02} demonstrate how such hot bubbles in CDM galaxies
%rapidly expand into the voids and avoid disturbing filaments. 
%\section{Results}

\begin{figure}
\hspace{0.13cm}\resizebox{4cm}{!}{\includegraphics{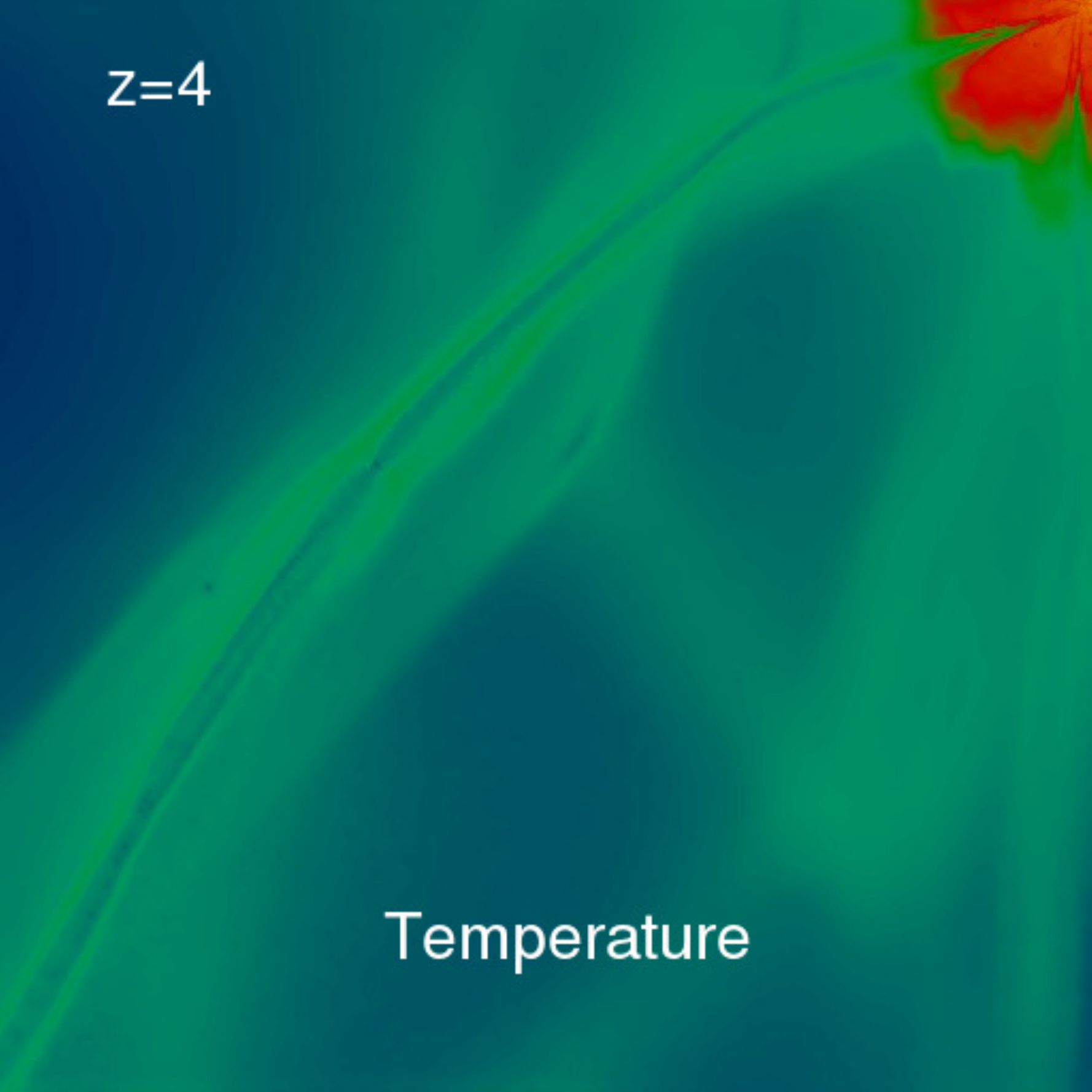}}
\hspace{0.13cm}\resizebox{4cm}{!}{\includegraphics{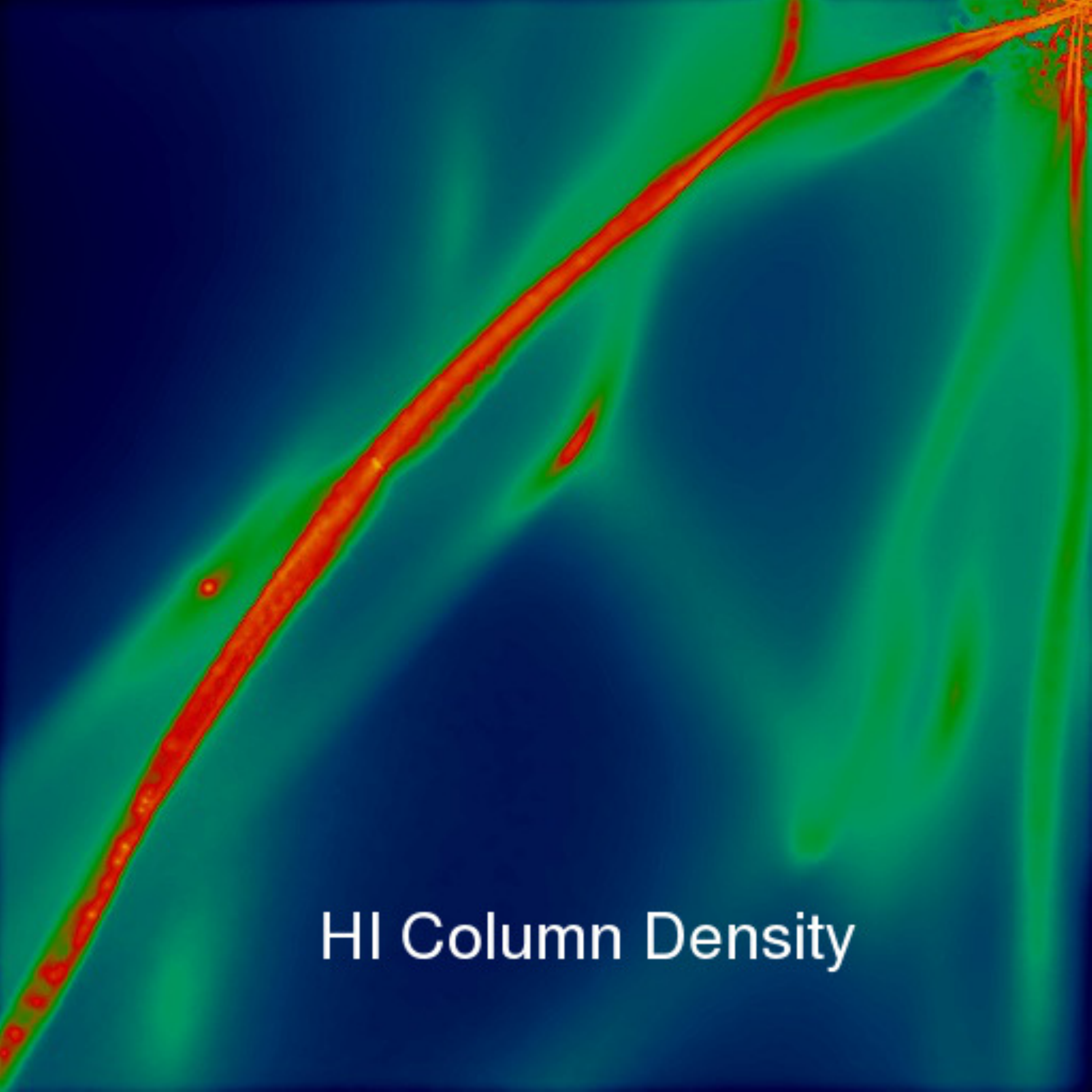}} 
\vspace{0.13cm}
\hspace{0.13cm}\resizebox{4cm}{!}{\includegraphics{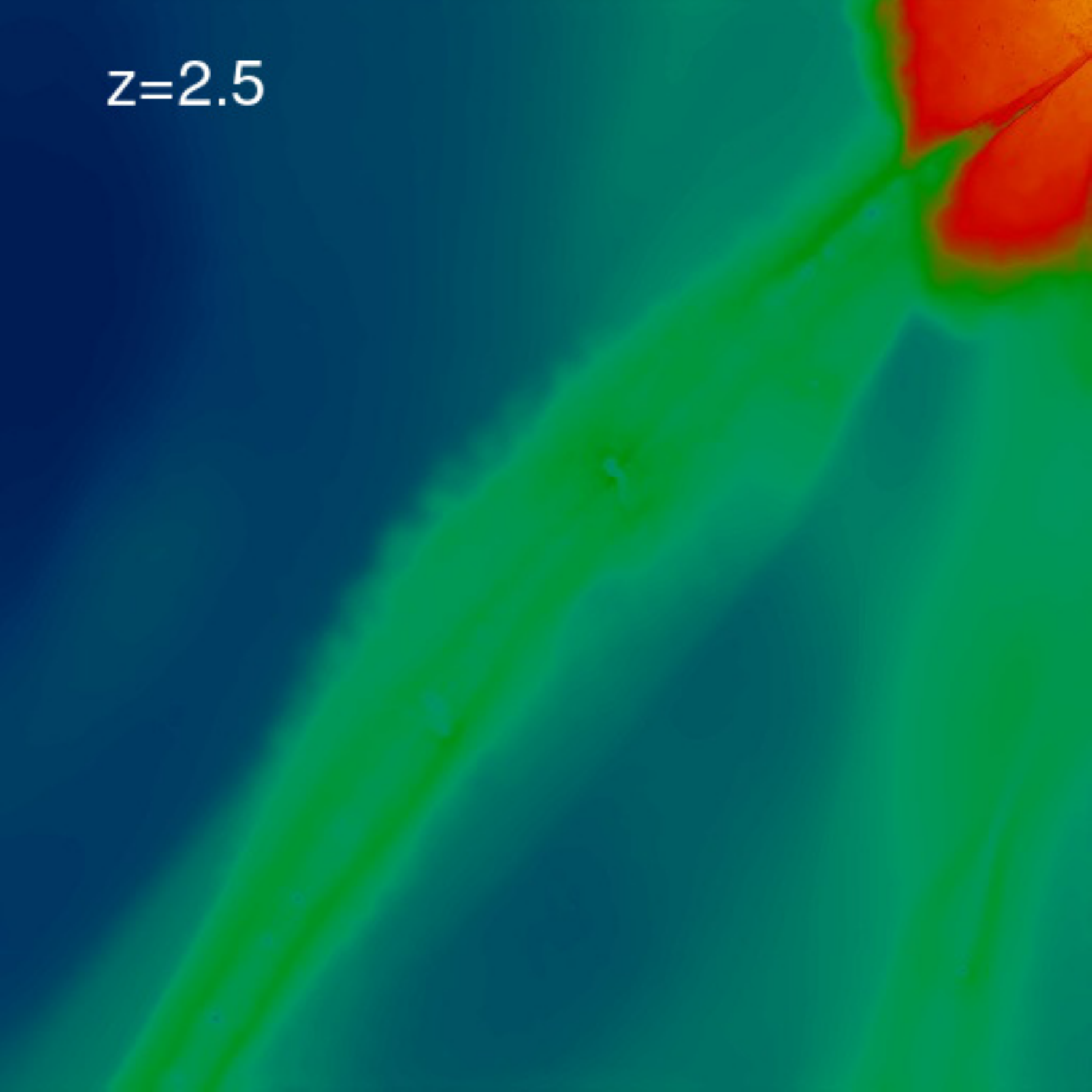}}
\hspace{0.13cm}\resizebox{4cm}{!}{\includegraphics{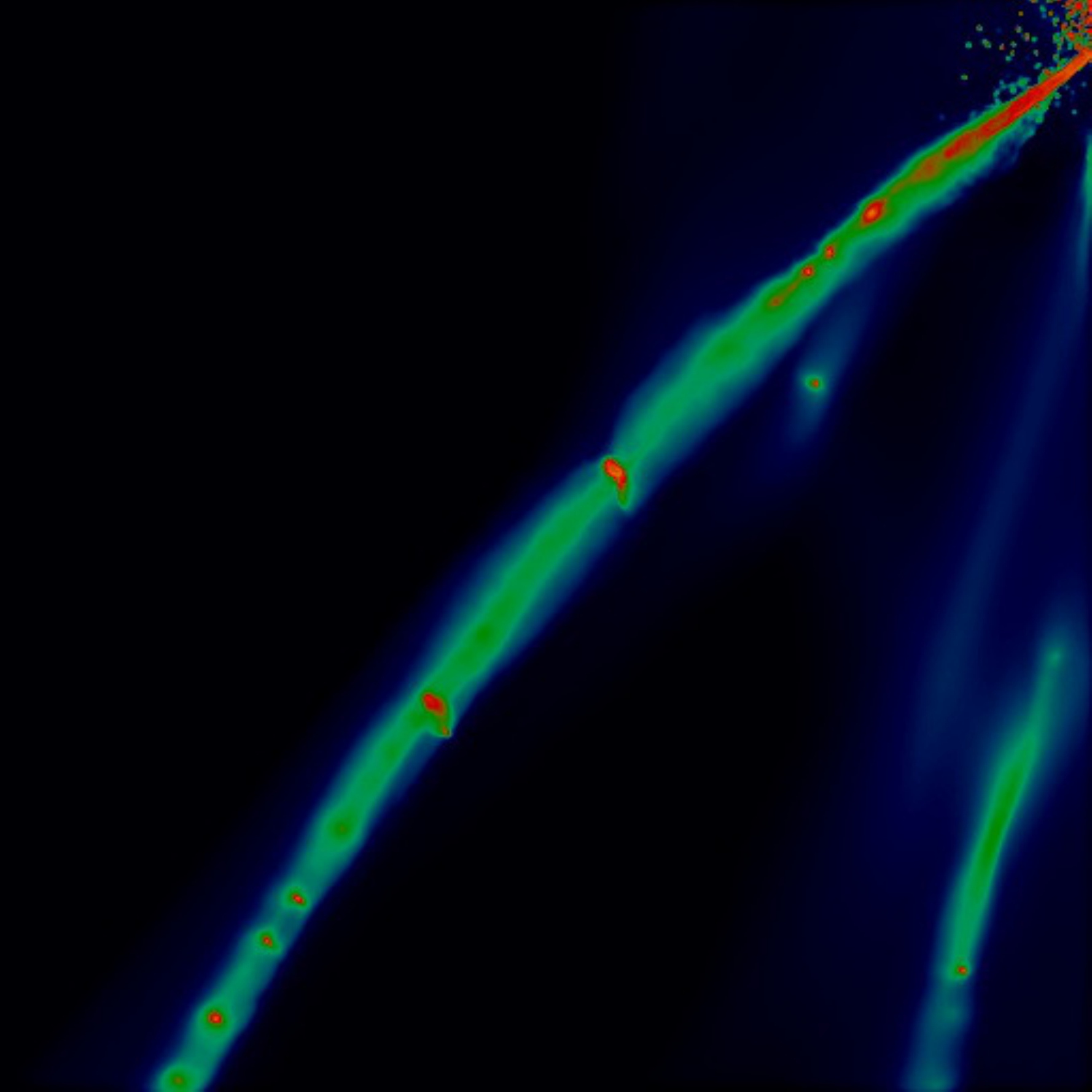}}
\hspace{0.13cm}\resizebox{4.2cm}{!}{\includegraphics{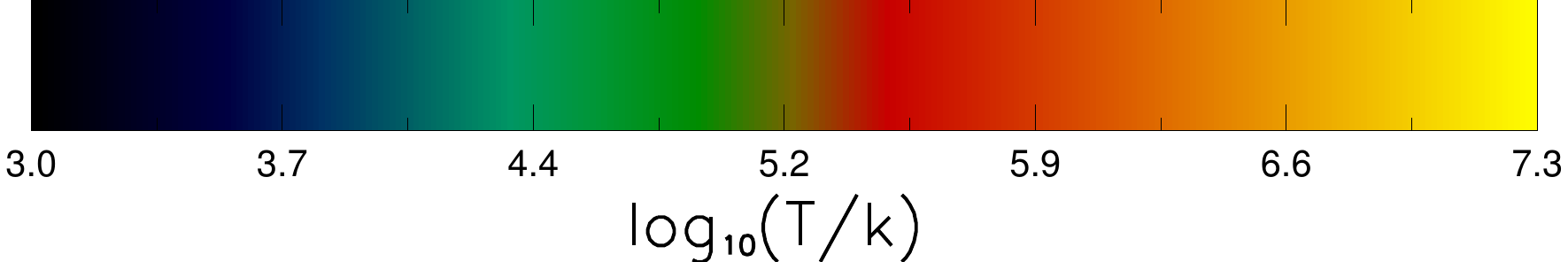}}
\hspace{0.13cm}\resizebox{4.2cm}{!}{\includegraphics{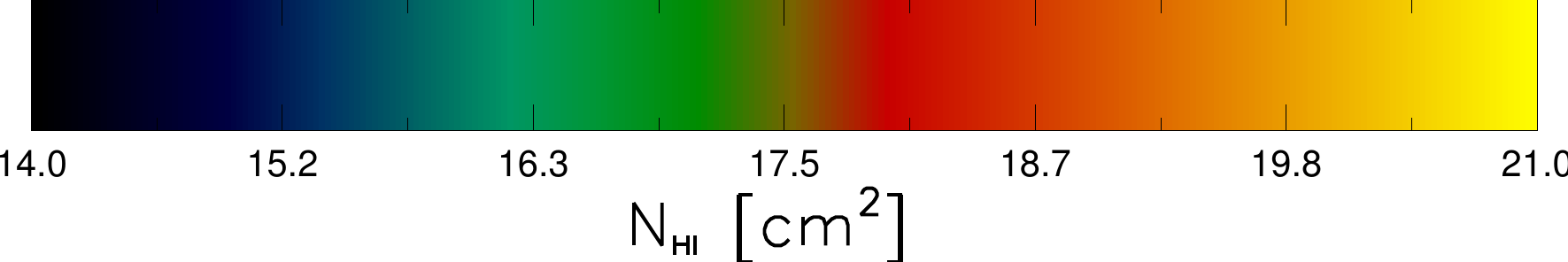}}
\caption{Zoomed-in temperature ({\em left panels}) and (total)
  hydrogen column density plots ({\rm right panels}) near the forming
  galaxy (the dense and hot structure on the top right of each panel)
  at $z=4$ ({\em top row}) and $z=2.5$ ({\em bottom row}). Each panel
  is 1.5$h^{-1}$ co-moving Mpc on a side. At $z=4$ the filament has
  very high column, $\gtsima 10^{20}$~cm$^{-2}$, and is so dense that
  the gas cools towards the centre. As this cold gas penetrates the
  hot halo of the forming galaxy, it remains mostly cool. The lower
  $z$ filament is wider and less dense, with some substructure that
  may be a consequence of artificial fragmentation of the underlying
  dark matter.}
\label{Fig2}
\end{figure}

\begin{figure}
\hspace{0.13cm}\resizebox{4cm}{!}{\includegraphics{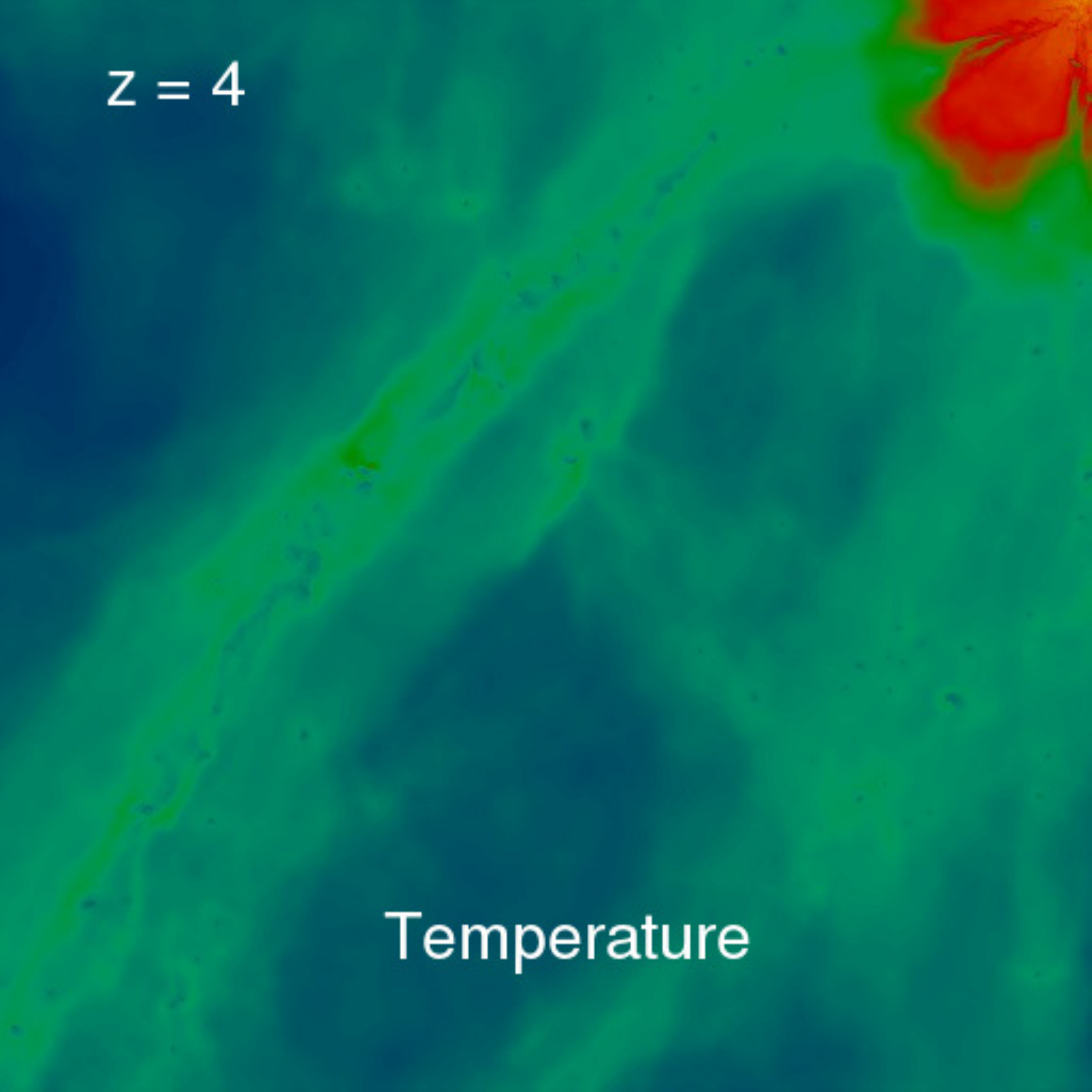}}
\hspace{0.13cm}\resizebox{4cm}{!}{\includegraphics{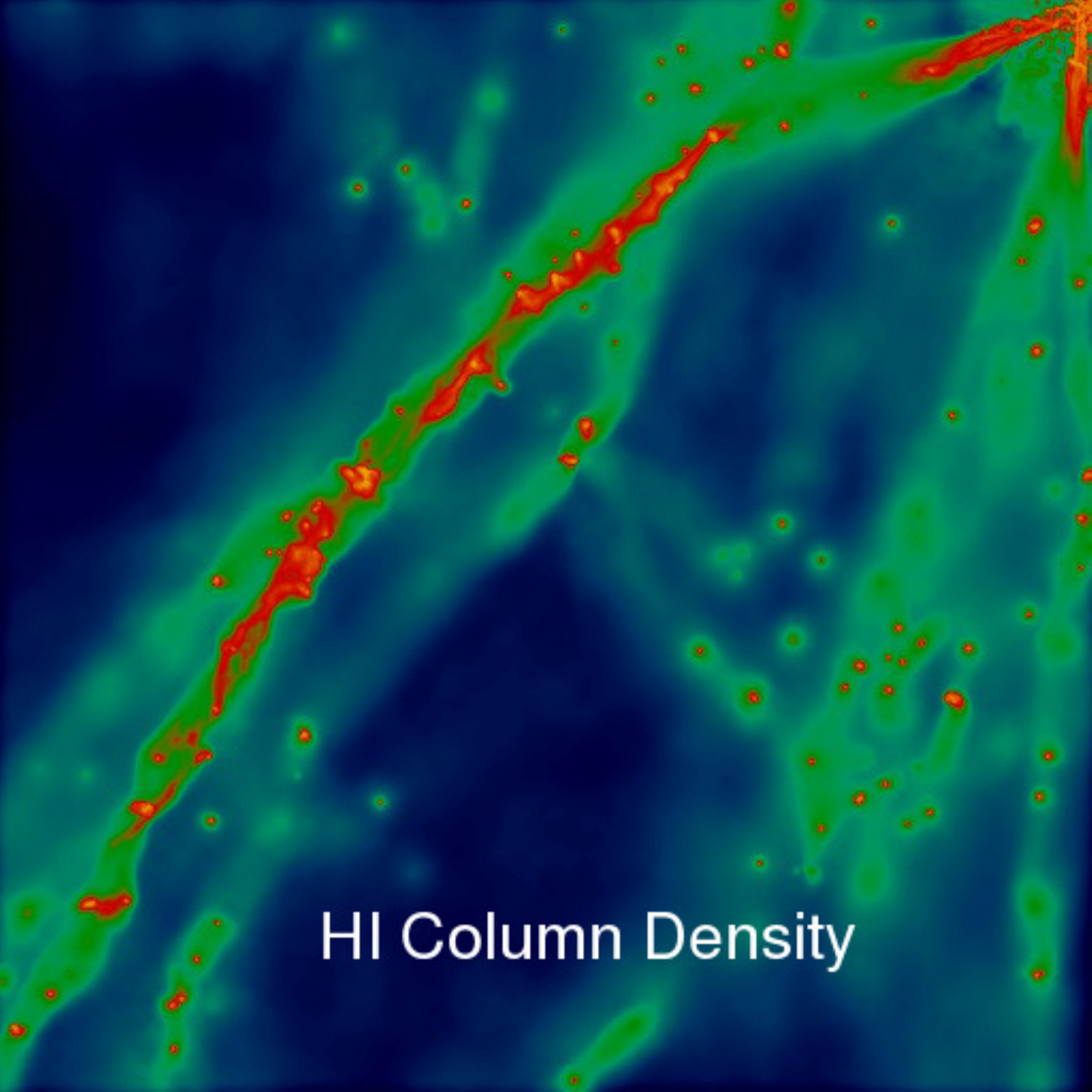}}
\vspace{0.13cm}
\hspace{0.13cm}\resizebox{4cm}{!}{\includegraphics{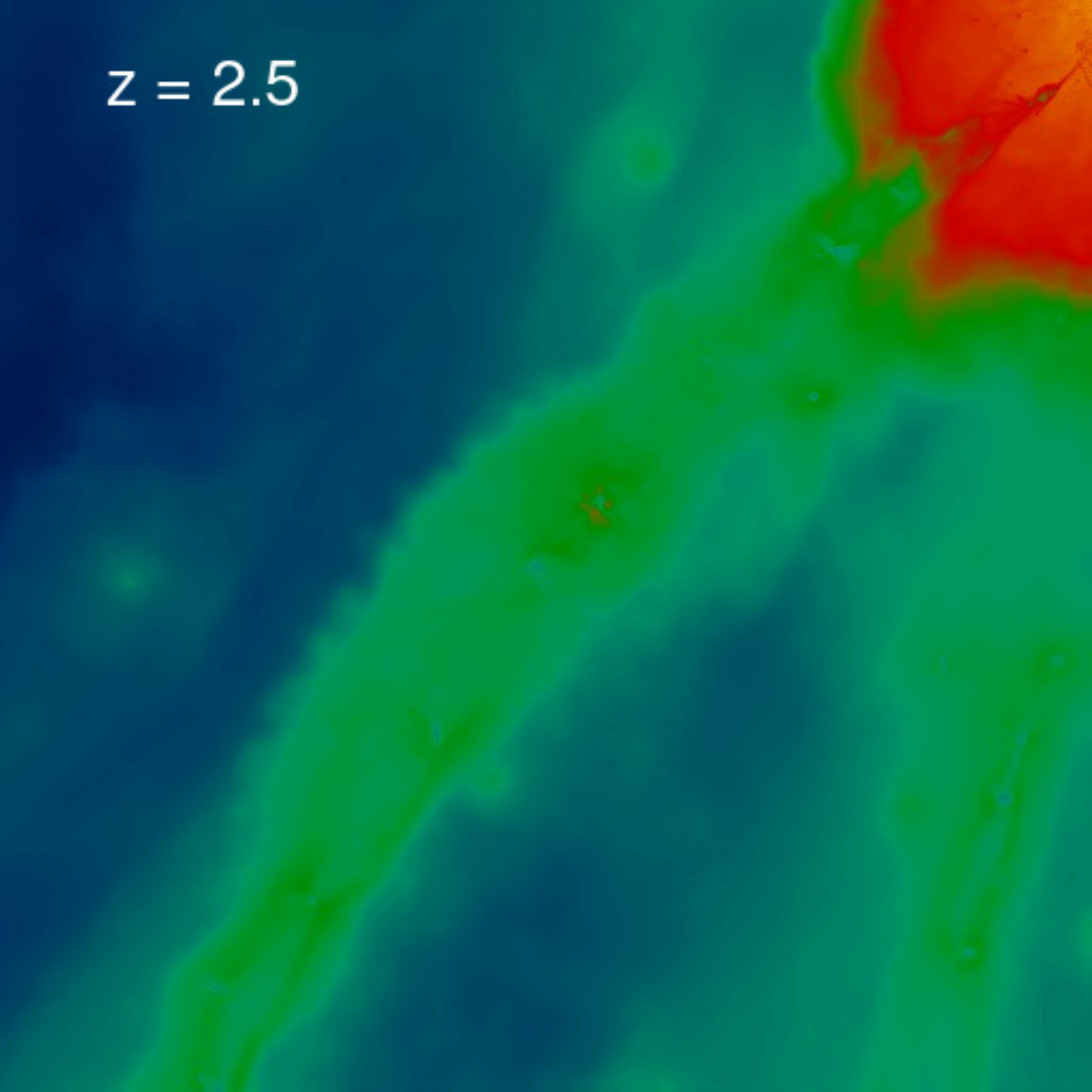}}
\hspace{0.13cm}\resizebox{4cm}{!}{\includegraphics{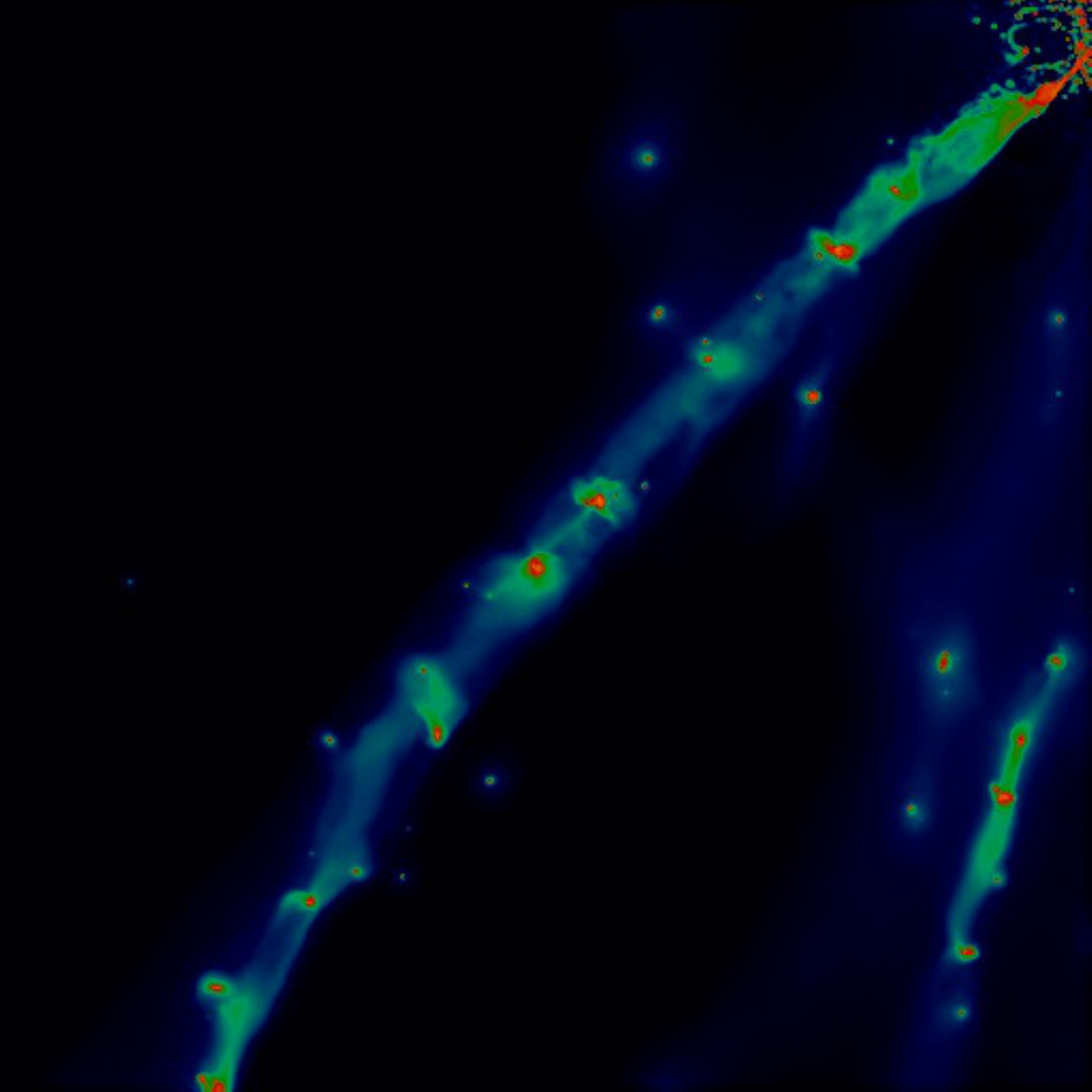}}
\hspace{0.13cm}\resizebox{4.2cm}{!}{\includegraphics{figs/tempbar1-eps-converted-to.pdf}}
\hspace{0.13cm}\resizebox{4.2cm}{!}{\includegraphics{figs/cor1-eps-converted-to.pdf}}
\caption{Temperature and density around the galaxy at redshift $z=4$
  ({\em top panels}) and $z=2.5$ ({\em bottom panels}) in the CDM
  case. Compare with Fig.2~ for the corresponding WDM case.}
\label{Figapp}
\end{figure}

A visual impression of the formation of a galaxy in WDM is presented
in Figure~1. Numerous smooth filaments dominate the $z>2.5$ dark
matter density field, each containing a considerable amount of gas. In
the highest redshift, $z=10$, snapshot (top row), the halo itself has
only barely started to form, yet the formation of the filaments is
already well advanced. Gas in the filaments is compressed, cools,
reaches the threshold for star formation, and is converted to stars,
producing the very filamentary \lq galaxy\rq\ seen in the top right
panel. Given our simplified star formation recipe, we find that there
is more star formation in filaments than in haloes up to redshift
$z=6$. At this stage, the simulated gas is assumed to become
reionized, and the photo-heating of the gas to temperatures $T\sim
10^4$~K decreases its density in filaments, suppressing star
formation.  Since we do not perform radiative transfer, the
photo-heating of filaments is overestimated as in reality they are
self-shielded, hence we may have underestimated how many stars would
form in the filaments.

At and below $z=4$ (bottom three rows), the formation of the halo
itself has caught up, leading to the appearance of a shock-heated halo
of gas, some of which cools to form the main galaxy - the more
familiar mode of galaxy formation in CDM. In the $z=2.5$ and $z=4$
snapshots, several high column density filaments ($N_{\rm
  H}>10^{17}-10^{20}$~cm$^{-2}$) extend over the full 4~$h^{-1}$~mega
parsecs of the panels, star formation is now mostly inside the main
halo, with remaining low levels of star formation in the filaments.
At $z=0$ (bottom row), dark matter filamentary structure around the
main halo is still very pronounced, and the halo and two main
filaments visible in the image contain hot, $T\sim 10^6$~K, gas. Stars
are overwhelmingly inside the main galaxy, with several smaller
satellites visible inside the filaments. Throughout the simulation,
artificial fragmentation of the dark matter in filaments
\citep{Wang07} is visible .

\begin{figure*}
\hspace{0.13cm}\resizebox{8cm}{!}{\includegraphics{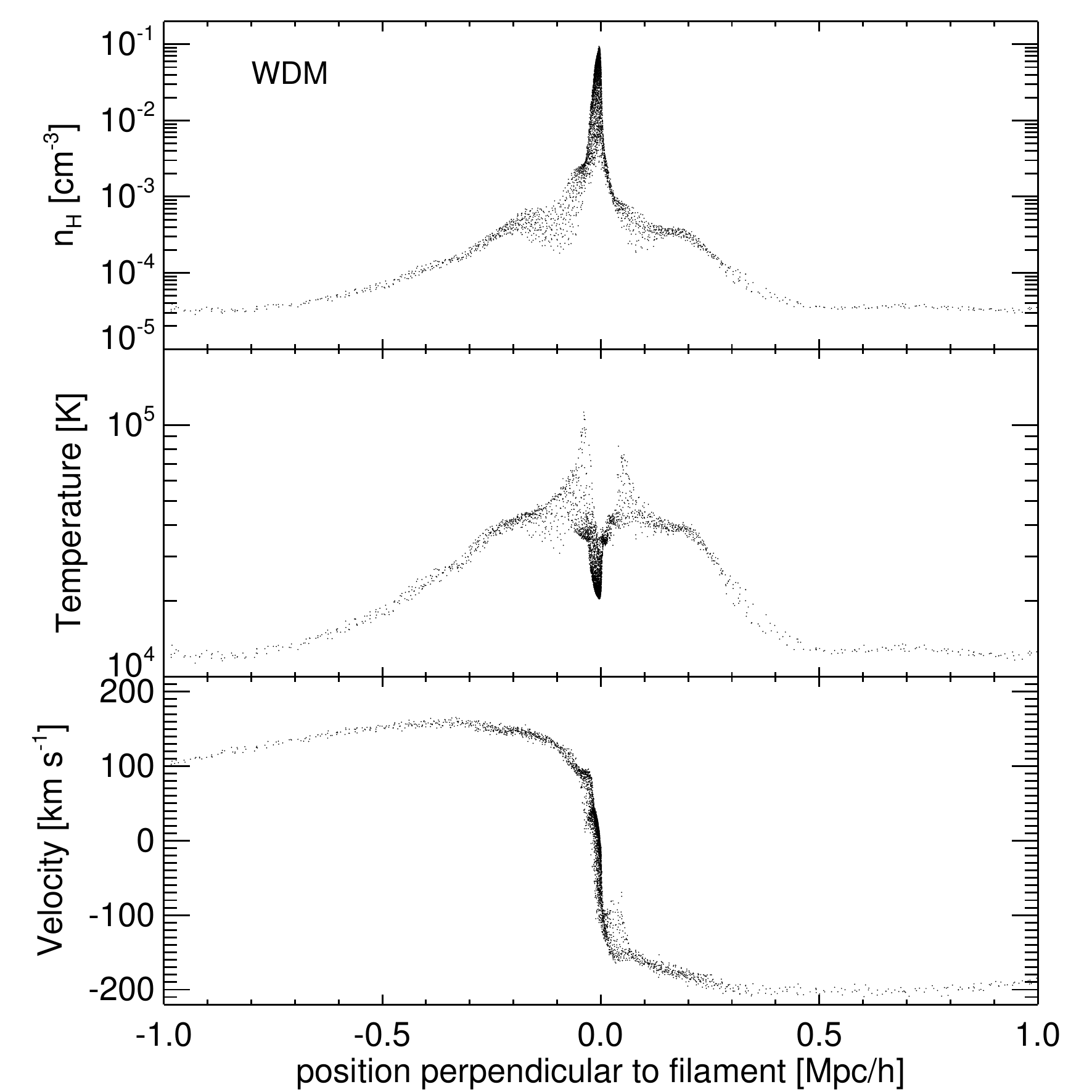}}
\hspace{0.13cm}\resizebox{8cm}{!}{\includegraphics{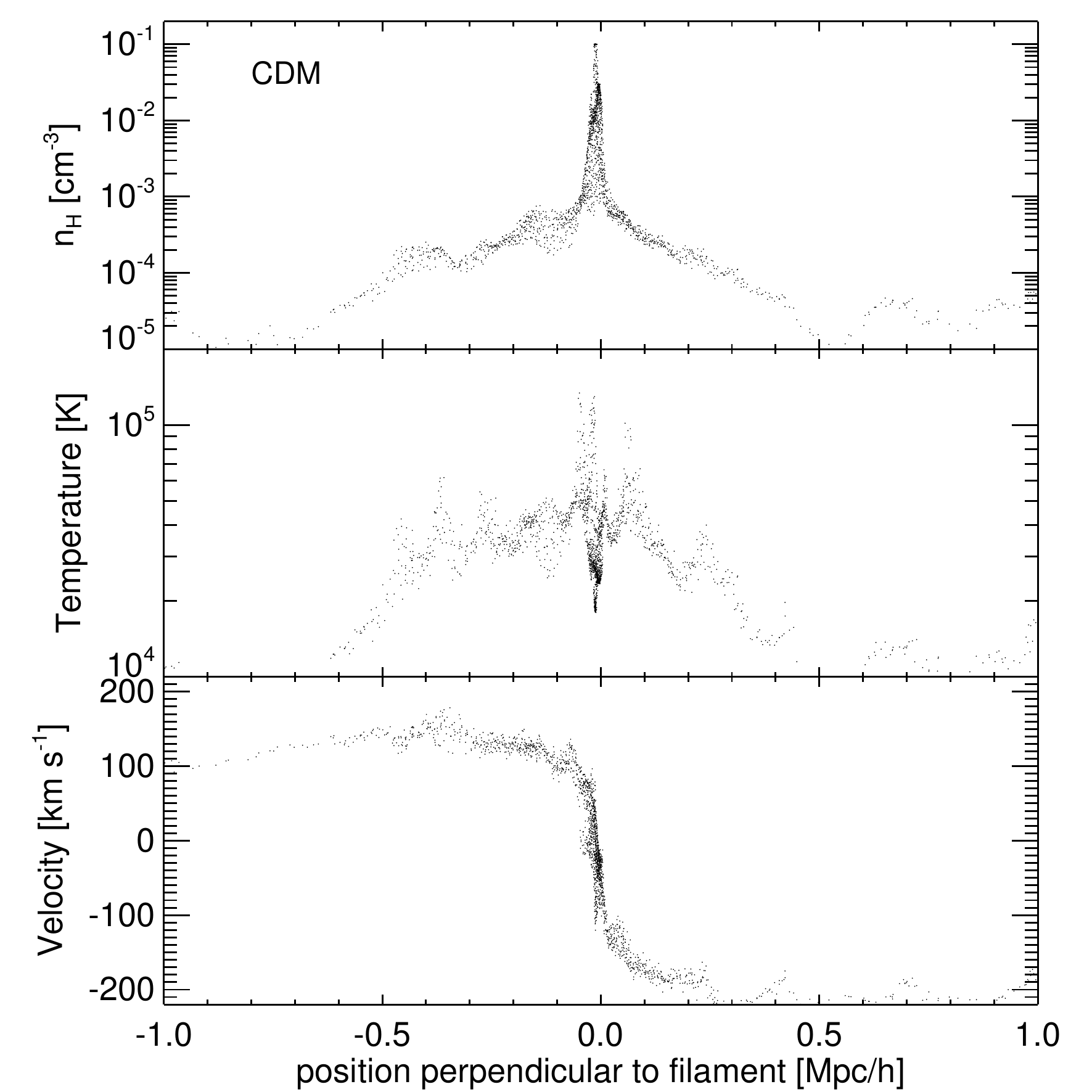}}
\caption{Left panel: Hydrogen gas number density, temperature
  and velocity of SPH particles ({\em top to bottom}) as function of
  distance to a WDM filament around a Mily Way-like progenitor galaxy
  at redshift $z=4$. Cold gas, $T\sim 10^4$~K, accretes onto the
  filament with velocities of $\sim 150$~km~s$^{-1}$, shocks to
  temperatures of $\sim 10^{4.5}-10^5$~K and densities $n_{\rm H}\sim
  5\times 10^{4}-10^{-3}$~cm$^{-3}$ at $100$~$h^{-1}$~kpc from the
  centre of the filament, cools radiatively and gets compressed to
  $T\sim 2\times 10^4$~K and $n_{\rm H}>0.1$~cm$^{-3}$. Above this
  density our star formation algorithm converts gas to stars. Right
  panel: same plot as the left panel but for the CDM simulation.}

\label{dots}
\end{figure*}

In Fig.~2 we zoom in closer to the galaxy in WDM. A very dense, cold filament
is seen to extent over the full 1.5$h^{-1}$ co-moving mega parsecs
extent of the plot. It is dense enough that when it penetrates the
halo of the forming galaxy (top right in each of the panels) it
remains cold, $T\sim 10^4$~K, near the base of the cooling function in
the absence of metals or molecules. The lower $z=2.5$ redshift
filament is thicker and slightly hotter, with hydrogen column $\sim
10^{17}$~cm$^{-2}$. It contains a number of higher column fragments,
which may result from artificial fragmentation in the underlying dark
matter. A direct comparison of the structure around the galaxy in CDM is possible 
by comparing to Fig.~3 in which we show the CDM version of the
  same plot. As expected, the filaments in CDM are much less smooth
  and break up in small clumps, an effect most striking at the higher
$z$,  where the WDM filaments are far denser and smoother.

We select a small section of a filament {\it in WDM} at $z=2.5$ of length $2h^{-1}$
co-moving kilo parsecs and rotate it so that its long axis lies along
the $x$-axis. We then plot  density, temperature and velocity of each
SPH particle in the segment against perpendicular distance from the
centre of the filament (left panel of Fig.~\ref{dots}). Gas
accretes onto this filament at speeds $\sim 150$~km~s$^{-1}$, shock
heats, then cools radiatively to the base of our cooling curve, as it
collapses to the centre. As it reaches a (hydrogen number) density of
$\gtsima 0.1$~cm$^{-3}$, we convert it to stars. Although one might
question the simple star formation criterion we employ here, or criticise the
fact that our filaments fragment artificially, the fact that gas
accretes onto WDM filaments and is able to cool to high densities
through atomic line cooling {\em before} accreting onto a halo, is
unambiguously demonstrated by Figs.1-4. 

In the right panel of Figure 4, we show a matched plot to the
left panel but for the CDM case. Comparing to WDM, the CDM filament
has more complex structure presumably arising from small scale
density fluctuation imprinted in the CDM initial condition. The gas
density histogram of the same set of SPH  particles is shown in
Figure 5. The WDM filament apparently has more SPH particles
distributed in the high density regime, consistent with the
appearance from the Figure 2 and 3.

\begin{figure}
\hspace{0.13cm}\resizebox{8cm}{!}{\includegraphics{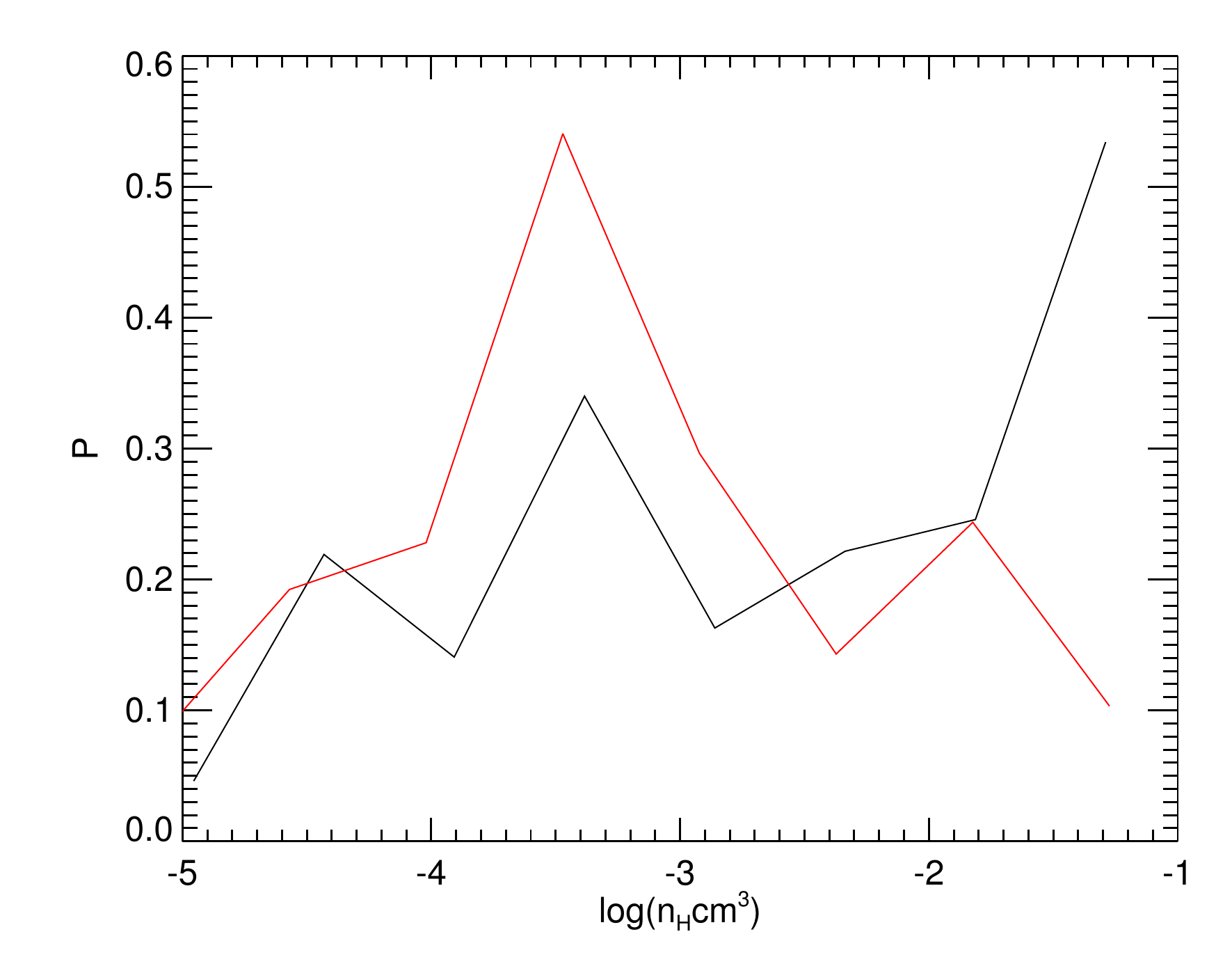}}
\caption{Comparison of gas density histogram of SPH particles shown
  in the figure~4. The red solid line is for the CDM simulation, while
  the black sold line is for the WDM run.}
\label{hist}
\end{figure}

\begin{figure*}
\hspace{0.13cm}\resizebox{8cm}{!}{\includegraphics{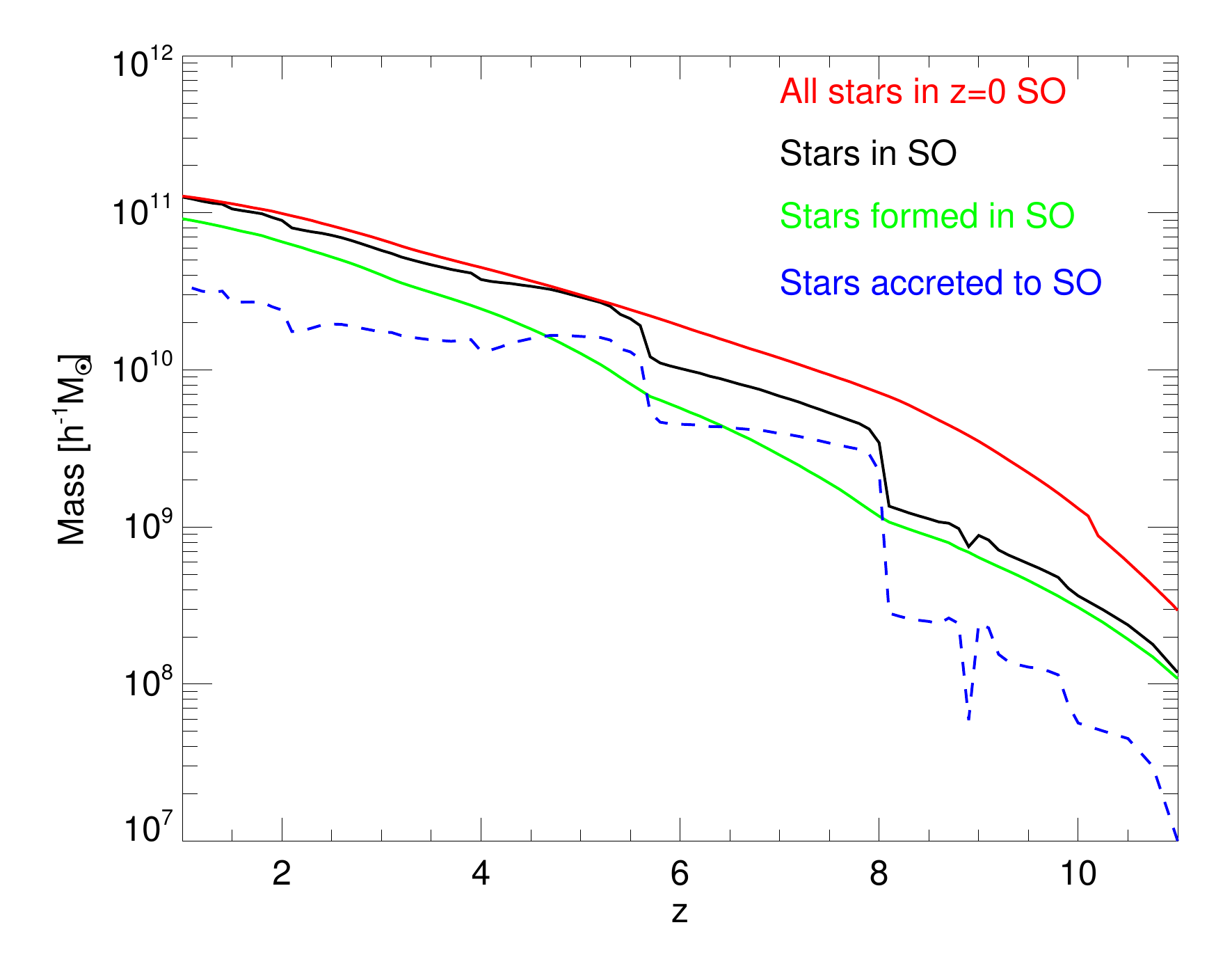}}
\hspace{0.13cm}\resizebox{8cm}{!}{\includegraphics{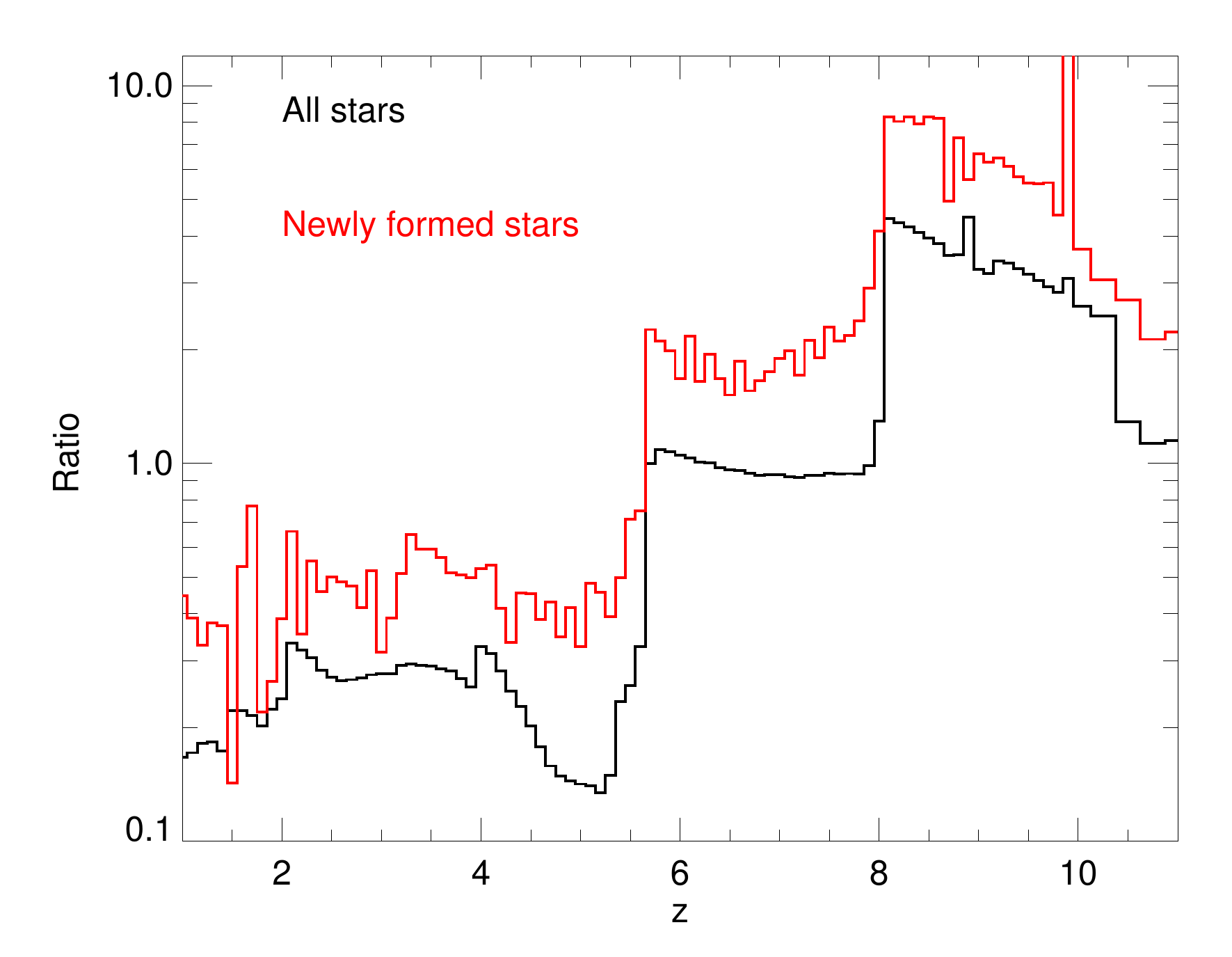}}
\caption{Star formation history of the Aq-A galaxy in warm dark
  matter. {\em Left panel}: the stellar mass formed up to redshift $z$
  that resides in the spherical over density ({\sc so}) galaxy at
  $z=0$ is shown as the {\em red line}, the stellar mass that already
  resides in the {\sc so}'s progenitor at redshift $z$ is shown as the
  {\em black line}. Above $z\sim 6$ the red line is far above the
  black line because {\em most} stars are in filaments. The stellar
  mass indicated by the black line is the sum of the green mass -
  stars formed in the {\sc so}, and the dashed blue line - stars
  accreted to the {\sc so}. {\em Right panel}: Ratio of all stars
  ({\em black line}) and newly formed stars ({\em red line}), outside
  the main halo to those in the main halo, as a function of
  redshift. The red line is above one for $z>6$, again showing that
  early on most stars form in filaments, and star formation in
  filaments continues to at least $z=1$. The black line is also mostly
  above one for $z>6$, meaning that most stars also reside in
  filaments at $z>6$.}
\end{figure*}

What is the fate of this gas? If the filament's column density is high
enough, $N_{\rm H{\sc I}}\gtsima 10^{17}$~cm$^{-2}$, it will
self-shield from the UV-background. Clearly the filaments around the
Aq-A WDM galaxy are (far) above this threshold. In our simulations,
this gas becomes self gravitating, and cools and collapses as far as
we allow it to. We speculate that if we were to include molecular
hydrogen formation, the gas would continue to cool further. The
situation is then very similar to that discussed by \cite{Gao07b} in
the context of first star formation in WDM.

Our simulated WDM filaments currently fragment artificially
therefore it would be unwise to over interpret their fate. In reality,
perturbations are seeded by the finite size of the filaments, i.e. by
the tidal field around them, because of the absence of small scale
structure in WDM. Perturbations started on these large scale can
propagate into the filament and potentially make them unstable. The
stability of gaseous filaments been investigated in the context of the
formation of cloud cores \cite[e.g.][]{Larson85, Inutsuka97}, and the
basic physics should still apply: if the growth rate of perturbations
is higher than the collapse rate of the filament as a whole, the gas
will fragment. Determining the growth rate requires much more detailed
simulations that include the chemistry of molecule formation and its
impact on cooling - and importantly avoid artificial fragmentation; we
leave this to future work.  Nevertheless we think it is likely that
filaments will fragment into clouds that eventually form stars at
sufficiently high redshift, $z\gtsima 2$ say for 1.5~keV thermal equivalent
WDM particles. At lower $z$, these stars accrete onto the main
galaxy. 

Although we have concentrated on star formation, the high column
density of the filaments in {\it WDM} above $z=2$ say, and their
remarkable uniformity and extent are very striking. Such structures
would be detectable in absorption as Lyman-limit (LLS) or Damped
Lyman-$\alpha$ systems (DLAs). It would be worth exploring if their
properties are sufficiently different from the LLS and DLAs that form
in CDM that we could distinguish the WDM-type models discussed here
from CDM. In particular, we expect the auto-correlation function of
LLS to be significantly different between CDM and WDM.

We can speculate about the fraction of stars that forms in filaments
in WDM versus in the main galaxy, but we caution that this is necessarily
uncertain given our simplified star formation implementation and even
more importantly our neglect of feedback. Traditionally the dark
matter and stellar distribution of a simulation is dissected in \lq
haloes\rq\ - defined by linking close enough particles together in a
friends-of-friends ({\sc fof}) halo \citep{Davis85} to select nearly
spherical regions within which the density is $\sim 200$ times the critical
density. In {\sc WDM} models this dissection does not work properly
because {\sc fof} will select not just the halo but most of the
surrounding filament as well. We therefore run a spherical over
density algorithm (hereafter denoted {\sc so}) centred at the location
of the most bound {\sc fof} particle to define the halo instead as a
{\em spherical} region within which the mean density is $\sim 200$
times the critical density. At high $z\gsim 7$ say, this assigns most
of the mass that eventually ends up in the $z=0$ object to filaments - consistent with the visual
appearance from Fig.~\ref{Fig1}.  We can now distinguish between stars
inside and outside the {\sc so}. We will refer to the most massive
{\sc so} as the progenitor of the $z=0$ galaxy.

How the $z=0$ galaxy is build-up of stars formed in its main
progenitor, and stars formed in filaments, is illustrated in Figure~5,
left panel. The red curve is the total mass of stars formed up to
redshift $z$, that are inside the {\sc so} galaxy at $z=0$. The black
curve is the mass of the {\sc so} galaxy at redshift $z$ - which is
much less than the red curve above $z=6$ because most stars are still
in filaments at this early time. The green and dashed-blue curve are
the stellar masses of the {\sc so} galaxy that formed {\em in situ}
(green), or were accreted onto the {\sc so} (dashed blue). The black
line in the right panel is the ratio of stellar mass outside the {\sc so}
galaxy to mass inside the {\sc so} galaxy, for stars that are in the
galaxy at $z=0$. The red line shows this ratio for stars that formed
recently - in the small redshift bins indicated in the panel. We see
that star formation is higher in filaments than in the {\sc so} galaxy
above $z=6$ - by up to a factor of 10. Two large accretion events (at
$z\sim 6$ and $z\sim 8$) drain filament stars into the {\sc so}
galaxy, and by $z\sim 6$ the mass in the {\sc so} galaxy becomes
comparable to the mass of stars in filaments.

Figure~6 quantitatively confirms the visual impression from Figs.~1-3,
that star and galaxy formation is dominated by filaments in the
WDM model, of this particular galaxy before $z=6$. Star
formation in filaments accounts for $\sim 3\times
10^{10}~h^{-1}M_\odot$ of stellar mass by $z=0$, or 15 per cent of the
$\sim 2\times 10^{11}h^{-1}M_\odot$ of the final galaxy. 

\section{Summary and discussion}

We have performed cosmological hydrodynamical simulations of the
formation of a Milky Way-like galaxy in a warm dark matter (WDM)
scenario. The model we investigate uses a CDM transfer function
exponentially cut-off below the free-streaming scale of a WDM particle
which is equivalent to that of a $1.5$~keV thermal relic. The simulation
includes radiative cooling from hydrogen and helium, inverse Compton
cooling off the CMB, and thermal bremsstrahlung, in the presence of an
imposed uniform optically thin UV/X-ray background, but ignores
cooling from metals and from molecules. The simulation uses a very
simple sub-grid model for star formation and neglects feedback from
star formation. We examined to what extent star formation in the dense
filaments that are characteristic for this WDM model contributes to
the final redshift $z=0$ stellar mass and its build-up.

At very high redshifts, $z\gsim 8$ say, dense and extended filaments
several co-moving mega parsecs long, form before dark matter haloes
themselves appear. Gas in those filaments is cold ($T\sim 10^4$~K, the
base of the cooling function in the absence of metals and molecules)
and dense ($n_{\rm H}\gsim 0.1$cm$^{-3}$) enough to form stars in our
sub-grid model for star formation. Filaments continue to dominate star
formation, with gas accreting onto them at high speed ($\sim
150$~km~s$^{-1}$), where it shock heats and subsequently cools
radiatively through atomic line cooling. The column density of gas
through these filaments is very high ($\gsim 10^{18}$~cm$^{-2}$), and
the presence of very long and narrow Lyman-limit systems (LLS) in WDM
is very striking. It might be possible to test observationally for the
presence or absence of such WDM filamentary LLS, for example by
studying the LLS correlation function, and hence constrain the nature
of the dark matter particle. Stars formed in filaments drain in
haloes. Reionisation - in these simulations assumed to occur at $z=6$
- causes the gas density in filaments to decrease, and star formation
in haloes starts to dominate; however stars continue to form in
filaments until $z\sim 2$, after which their column density drop below
the threshold for star formation. By $z=0$, 15 per cent of stars in
the final galaxy formed in filaments. 

We stressed that there is no accepted theory of how stars form in
filaments, nor has it been investigated if and how star formation in
filaments is regulated by feedback. In addition, our simulation
suffers from the well known artificial fragmentation of the dark
matter. Consequently we are hesitant to draw quantitative conclusions
from this run. However what the simulation demonstrates unambiguously
is that in this WDM model, very long and dense filaments form around
Milky Way-like proto-galaxies, which would be observable as LLS or
DLAs. Gas in these filaments cools radiatively through atomic line
cooling, and can shield itself from the UV-background. The thermal
instability that results from this is likely to lead to star
formation, even at lower redshifts $z\sim 2$. In the simulation, this
results in the appearance of curiously shaped stringy \lq chain\rq\
galaxies, which, if observed, might be a strong indication that the
dark matter is warm.

\section*{acknowledgments}
The simulations used in this work were carried out on the Lenova
Deepcomp7000 supercomputer of the super Computing Centre of Chinese
Academy of Sciences, Beijing, China. LG acknowledges support from the
{\small NSFC} grant (Nos. 11133003,11425312), the Strategic Priority Research
Program ''The Emergence of Cosmological Structure''  of the Chinese
Academy of Sciences (No. XDB09000000), {\small MPG} partner Group
family, and an {\small STFC} Advanced Fellowship, as well as the
hospitality of the Institute for Computational Cosmology at Durham
University.  VS acknowledges support by the European Research Council
under ERC-StG grant EXAGAL-308037.This work was supported by the
Science and Technology Facilities Council [grant number ST/F001166/1]
and by the Interuniversity Attraction Poles Programme initiated by the
Belgian Science Policy Office ([AP P7/08 CHARM]). It used the DiRAC
Data Centric system at Durham University, operated by the Institute
for Computational Cosmology on behalf of the STFC DiRAC HPC Facility
(www.dirac.ac.uk). This equipment was funded by BIS National
E-infrastructure capital grant ST/K00042X/1, STFC capital grant
ST/H008519/1, STFC DiRAC Operations grant ST/K003267/1 and Durham
University. DiRAC is part of the National E-Infrastructure. The data
used in the work is available through collaboration with the
authors.

\bibliographystyle{mnras}
\bibliography{paper}

\section{Appendix}
\begin{figure}
\hspace{0.13cm}\resizebox{8cm}{!}{\includegraphics{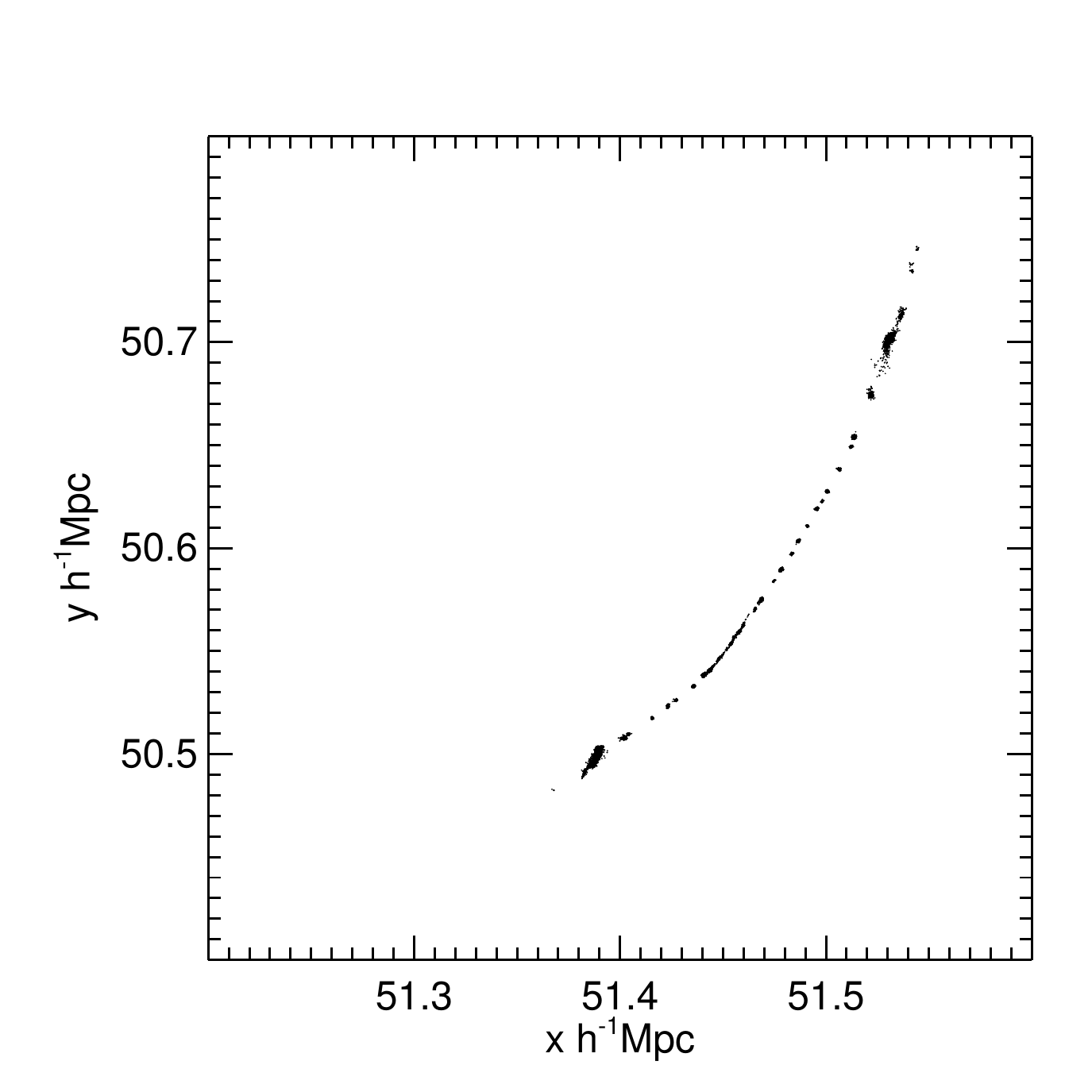}}
\caption{Positions of star particles in a filament of our new
  simulation at z=10.}
\label{figz10}
\end{figure}

We have re-run our WDM simulation to redshift $z=6$ by adopting a higher
density threshold $n_h = 5/cm^3$ for the star formation. The result is
qualitatively similar to our previous one, namely star formation still
occur in WDM filaments. Of course the exact amount of stars formed at
given epoch is lower as we expected. As an example, in
Figure~\ref{figz10}, we show positions of all star particles formed in a
filament of our new simulation at z=10.

\end{document}